# STARDUST FINDINGS – IMPLICATIONS FOR PANSPERMIA

**Pushkar Ganesh Vaidya**

Indian Astrobiology Research Centre (IARC)

P O Box 8482, Mandpeshwar, Borivli (West), Mumbai – 400103, Maharashtra, India

+ 91 9870798093 / pushkar@astrobiology.co.in

**ABSTRACT**

In January 2004, the Stardust spacecraft flew through the dust of comet 81P/Wild 2 and captured specks of the cometary dust. On analysis of the comet 81P/Wild 2 samples it was found that they contain materials formed in the coldest and hottest regions of the early solar nebula, strongly suggesting 'mixing' on the grandest scale. Here, it is suggested that if microorganisms were present in the early solar nebula, as required by the hypothesis of cometary panspermia, then in the light of the Stardust findings, life was already present in the very material that formed planetary bodies.

## 1. COMETARY PANSPERMIA

The hypothesis of cometary panspermia needs a small fraction of microorganisms present in the interstellar cloud from which the solar system formed to have retained viability, or to be capable of revival after being incorporated into newly formed comets. Some comets owing to orbital disruptions get deflected towards the inner solar system thus carrying microorganisms onto the Earth and other inner planets. So, as per cometary panspermia life was first brought to Earth, about 4 billion years ago by comets and they continue to do so (*N.C. Wickramasinghe et al.* 2003). The hypothesis of cometary panspermia is not yet vindicated.

## 2. COMETS

Comets formed in the early stages of the condensation of the solar system. They contain the most pristine material available from that epoch which helps us understand conditions that existed in the young solar nebula more than 4.6 billion years ago. The major comet reservoirs are Oort cloud and Kuiper belt as well as Trans-Neptunian scattered disc. Comets spend most of their lives in these reservoirs (*Crovisier, J.* 2006).

The aging or evolutionary effects that a comet nucleus will experience can be divided into four primary areas: the *precometary phase*, where the interstellar material is altered prior to incorporation into the nucleus; the *accretion phase,* the period of nucleus formation; the *cold storage phase*, where the comet is stored for long periods at large distances from the Sun; and the *active phase* where the comet undergoes drastic changes owing to increased solar insolation as it approaches the inner solar system (*Meech K.*1999) and (*Meech K. J. & Svoren J.* 2005).

## 3. STARDUST FINDINGS

In January 2004, the Stardust spacecraft flew through the dust of comet 81P/Wild 2 and captured specks of the cometary dust.

On analysis of the comet 81P/Wild 2 samples, materials like Olivine, Calcium Aluminum Inclusions (CAIs) (*Sandford et al.* 2006) which formed at extremely high temperatures and Polycyclic Aromatic Hydrocarbons (PAHs) which formed at very low temperatures were found (*Sandford et al.* 2008).

Stardust findings made it clear that some cometary materials formed in regions with temperatures above 2000 K while others, especially the ice components, appear to have been formed in regions below 40 K, only a few degrees above absolute zero (*Brownlee et al*. 2006).

## 4. CONCLUSIONS - IMPLICATIONS FOR PANSPERMIA

One of the cornerstones of the hypothesis of cometary panspermia is the requirement of the presence of microorganisms in the solar nebula.

Stardust findings strongly suggest that *precometary phase* included mixing on the grandest scales between the coldest and hottest regions of the solar nebula. So, if microorganisms were indeed present in the solar nebula then they got well 'mixed up' and not only got incorporated into comets but rather into every body of the solar system.

We do not know if microorganisms could have survived the turbulent 'mixing' of the early solar nebula and later the violent planetary formation processes. However if they somehow did, then it means life was already present in the very material that formed the planetary bodies. This conjecture is testable in future when various planetary bodies are studied in great detail.